# Insights into Complex Brain Functions Related to Schizophrenia Disorder through Causal Network Analysis

Akram Yazdani[1]*, Raul Mendez Giraldez[2], Ahmad Samiei[3,4]

[1] Department of Neuroscience, Icahn School of Medicine at Mount Sinai, New York, NY, USA
[2] Lineberger Comprehensive Cancer Center, University of North Carolina School of Medicine, Chapel Hill, NC, USA
[3] Climax Data Pattern, Houston, Texas, USA
[4] Hasso Plattner Institute, 14482 Potsdam, Germany

*To whom correspondence should be addressed

**Abstract**

Gene expression represents a fundamental interface between genes and environment in the development and ongoing plasticity of the human organism. Individual differences in gene expression are likely to underpin much of human diversity, including psychiatric illness. Gene expression shows a distinct regulatory pattern in different tissues. Therefore, brain tissue analysis provides insights into brain disorder mechanisms. Furthermore, mechanistic understanding of gene regulatory pattern can be provided through studying the underlying relationships as a complex network. Identification of brain specific gene relationships provides a complementary framework in which to tackle the complex dysregulations that occur in neuropsychiatric and other neurological disorders.

Using a systems approach established in Mendelian randomization and Bayesian Network, we integrated genetic and transcriptomic data from the common-mind consortium and identified transcriptomic causal networks in observational studies. Focusing on Schizophrenia disorder, we identified high impact genes and revealed their underlying pathways in brain tissue. In addition, we generated novel hypotheses including genes as causes of the schizophrenia-associated genes and new genes associated with Schizophrenia. This approach may facilitate a better understanding of the disease mechanism that is complementary to molecular experimental studies especially for complex systems and large-scale data sets.

**Introduction**

Due to physical context of tissue environment, cells from different tissues exhibit unique functions that define tissue specific phenotype. These specific process and common processes among all cells that are essential for survival are ultimately controlled by gene regulatory networks. Gene regulatory networks alter genes expression and control the extent of that expression. Therefore, focusing on brain tissue analysis provides specific understanding of activity and regulation of multiple systems in brain and consequently dysregulation of genes in neurological disorder, such as schizophrenia.

Schizophrenia (SCZ) as a heritable complex mental disorder is a severe psychiatric disorder with a world-wide prevalence of 1% that is characterized by abnormalities in thought and cognition. Recently, multiple genetic studies, including linkage scans and their meta-analyses, candidate gene association analyses, differential expression analyses and genome-wide association studies (GWAS) have investigated genes/markers and chromosomal regions for SCZ (Ng et al., 2009; International Schizophrenia Consortium et al., 2009; Shi et al., 2009; Stefansson et al., 2009; Ripke et al., 2014). These studies show that SCZ disorder involves changes in multiple genes, each conferring small and incremental risk that potentially converge in deregulated biological pathways, cellular functions and local circuit changes, eventually scaling up to brain region pathophysiology (Belmaker and Agam, 2008; Sibille and French, 2013). In addition, correlation based networks have been applied on schizophrenia candidate genes (He, Chen and Evans, 2007; Wright et al., 1999; Bullmore and Sporns, 2009; Fromer et al., 2016). Such networks tend to result in many dependencies that arise from indirect interactions in the underlying anatomy and face limited success in finding functional genes and intervention targets (Sullivan, Kendler and Neale, 2003).

Utilizing systems approaches established in recognition of hierarchal structure of many systems biology can provide conceptual clarity and biological accuracy to better understand disease mechanism. These applications narrow search space down to a subset of biological components with high impact on disease end points and identify underlying pathways and intervention targets with reproducible results (Yazdani et al., 2018). Since the connections among genes are disturbed in the patients with SCZ disorder, manipulating in intervention targets may reconnect the genes. These features facilitate a mechanistic understanding/causal interpretation (Yazdani et al., 2015; Yazdani et al., 2014; Pearl, 2000) and make systems approaches complementary to molecular experimental studies especially for complex systems and large-scale data sets.

We here focused on integrating genomics and transcriptomics and utilized a systems approach called genome directed acyclic graph (the G-DAG algorithm) established in Mendelian Randomization and Bayesian graphical modeling (Yazdani *et al.*, 2016a) with reproduced findings (Yazdani et al., 2018). The RNA sequence data were from post-mortem dorsolateral prefrontal cortex which controls complex, higher-level cognitive and executive functions. We modified the G-DAG algorithm and built a transcriptomic causal network which reveals the underling transcriptomic pathways. Across the network, we identified modules and individual genes with high impact in the system. Our examination of the transcriptomic network revealed that each gene pathway includes a small subset of genes in the system. Mapping the SCZ associated genes on the transcriptomic network, we identified modules related to SCZ and hypothesized the genes that are causes of the SCZ associated genes and consequently are intervention targets. For instance, the gene *MAS1* is identified as the cause of SCZ associated gene *NRXN3*. In addition, we hypothesized new genes associated with SCZ, such as *GRIN3A*, as well as new targets for intervention, such as COMT. Moreover, identification of underlying transcriptomic pathways reveals how the effect of the intervention spread in the system.

**An overview on application of Mendelian randomization/instrumental variables**

To gain sufficient understanding and predict behavior of a system, we utilize Mendelian randomization technique using genetic variants as instrumental variables, which recently has gained attentions in biomedical applications. Instrumental variable approaches control for unmeasured confounding relative to other approaches such as regression, matching, and propensity score methods (Dawid, 2007; Sheehan *et al.*, 2008; Yazdani et al., 2016c). This feature leads to robust results with mechanistic understanding (i.e. causal interpretation) even in the presence of unmeasured confounders and reverse causation if the utilized genetic variants qualify as instrumental variables:

1. Being associated with a transcript of interest (exposure) that is a potential cause of an outcome transcript.
2. Not being associated with any confounders (measured and unmeasured) of the two transcripts of interest
3. Not being associated with the transcript outcome directly but only through the transcript of interest.

These conditions are visualized in Figure 1 in a small scale.

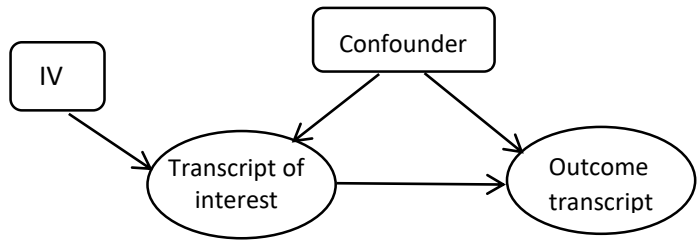

Figure 1. Visualization of instrumental variable assumptions. IV stands for instrumental variable.

Genotype features such as pleiotropy, the presence of linkage disequilibrium, genetic heterogeneity, lack of knowledge about the confounders, and population stratification, may violate the assumptions of instrumental variable and bring limitations to application of genetic variants (Sheehan *et al.*, 2008). Although the assumptions must be justified from background knowledge of the underlying biology and none of them are statistically testable, we can take some steps toward holding these assumptions. Evaluating the assumptions adds to the credibility of the analyses and there are some methods to this end.

Some approaches aim to find a single genetic variant strongly correlated with the variable of interest. These approaches are limited to identify sufficient numbers of instrumental variables for a large-scale data set (Inouye *et al.*, 2010) and possibility violate some of the assumptions due to the genotype features such as pleiotropic effect. Here, we focus on genome directed acyclic graph (G-DAG) algorithm that aims to hold the underlying assumptions through three main features as following:

1. Extracting information from several single nucleotide polymorphism (SNP)/genes to create stronger instrumental variables than a single SNP/gene.

2. Possibility of allocating multiple instrumental variables to a transcript to explain the variation of the transcript sufficiently.

3. Generating independent instrumental variables to avoid violation of the assumptions, due to genetic variants' properties such as pleiotropic effect.

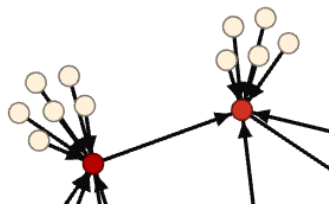

Figure 2. Visualization of some features of the G-DAG algorithm. White: generated instrumental variables from multiple SNPs. Brown: variables of interest here transcripts. The aim is identification of underlying relationship between the transcripts (browns) using the instrumental variables (whites).

These features of the G-DAG algorithm provide opportunities to explain the variation in transcripts that are due to genetic variants and free of confounding.

**Materials and methods**

**Study sample:** The RNA sequence (RNA-seq) data are from post-mortem dorsolateral prefrontal cortex which controls complex, higher-level cognitive and executive functions. These data are available in CommonMind Consortium (CMC; http://www.synapse.org/CMC). Following data normalization, 16,423 genes (based on Ensembl models) were expressed at levels sufficient for analysis, of which 14,222 were protein-encoding. Validation using PCR showed high correlation (r > 0.5) with normalized expression from RNA-seq for the majority of genes assessed. Covariates for RNA integrity (RIN), library batch, institution (brain bank), diagnosis, age at death, genetic ancestry, post-mortem interval and sex together explained a substantial fraction (0.42) of the average variance of gene expression and were thus employed to adjust the data for the analyses (Fromer *et al.*, 2016).

Samples were genotyped on the Illumina Infinium HumanOmniExpressExome array (958,178 single-nucleotide polymorphisms, or SNPs). These genotypes were used to detect SNPs that have an effect on gene expression, to estimate ancestry of the samples and to ensure sample identity across DNA and RNA experiments. The distribution of ethnicities shows Caucasian 80.7%, African-American 14.7%, Hispanic 7.7%, East Asian 0.6%. To prevents the study be confounded due to different ethnicities, we included 80.7% Caucasian (209 samples) in this study.

**Methods**

To investigate complex networks among transcripts utilizing Mendelian randomization technique, we first generated instrumental variables from genetic variants across the genome. Through this approach, only a few instrumental variables significantly were found related to the transcripts. The result was consistent with the fact that gene transcription is strictly controlled by the interplay of regulatory events at gene promoters and gene-distal regulatory elements called enhancers. Therefore, we concluded that gene expression is affected by genetic variations nearby genes (orientation, position, and distance) (Yazdani et al., 2017). This result is also reported recently by (Aguet *et al.*, 2017). To generate the IVs based on genetic variants nearby genes, we included variants in 40 kb upstream of transcription start site (TSS) and 40kb downstream of transcription end site (TES) to involve variants in genes and their promoters. In addition, we used high-resolution 3D maps of chromatin (Hi-C) (Won *et al.*, 2016) for reflecting the secondary structure properties of looped DNA within a nucleus. By selecting those variants, we aimed to consider the variants that mediate the effects of cis-regulatory elements via both short- and long range interactions.

To generate instrumental variables and fulfill the underlying assumptions, we applied multiple correspondence analysis (Ng *et al.*, 2009) over a set of variants nearby each gene. The method multiple correspondence analysis, which is a part of a family of descriptive methods, is the multivariate extension of correspondence analysis to analyze tables containing three or more variables. In addition, multiple correspondence analysis is considered as a generalization of principal component analysis for categorical variables and allows to investigate the pattern of relationships of several categorical variables in complex data sets. Despite principal component analysis, multiple correspondence analysis is not limited to linear relationship among genetic variants and has the ability to identify their non-linear relationship for generating instrumental variables.

After generating instrumental variables, we applied the G-DAG algorithm to build transcriptomic network and employed Hamming distance (Tsamardinos, Brown and Aliferis, 2006) to assess quality of the fit in identified network. The G-DAG algorithm selects the generated instrumental variables that have significant relationship with the transcripts. The instrumental variables help identify robust directions across Bayesian networks that are not identified through interaction among genes (Yazdani *et al.*, 2016a). The stability of the identified Bayesian networks in observational study is established in Mendelian principles since the genetic inherited variation is the cause of phenotypic variation (here gene expression). Therefore, these stable Bayesian networks are called causal networks in observational studies (Pearl, 2000; Sheehan *et al.*, 2008; Yazdani et al., 2016c).

**Results**

We generated 4,641 IVs from 958,178 SNPs and identified the transcriptomic network over a set of 1,181 transcripts. These transcripts are belongs to a class of genes with highest number of SCZ associated gene in an study by Fromer *et al.* (2016). We mapped the SCZ associated genes on the identified transcriptomic network with blue colors presented in Figure 3. The identified transcriptomic network revealed underlying relationships (pathways) among genes. To extract information from the transcriptomic network, we measured the network properties, including high impact genes and modules.

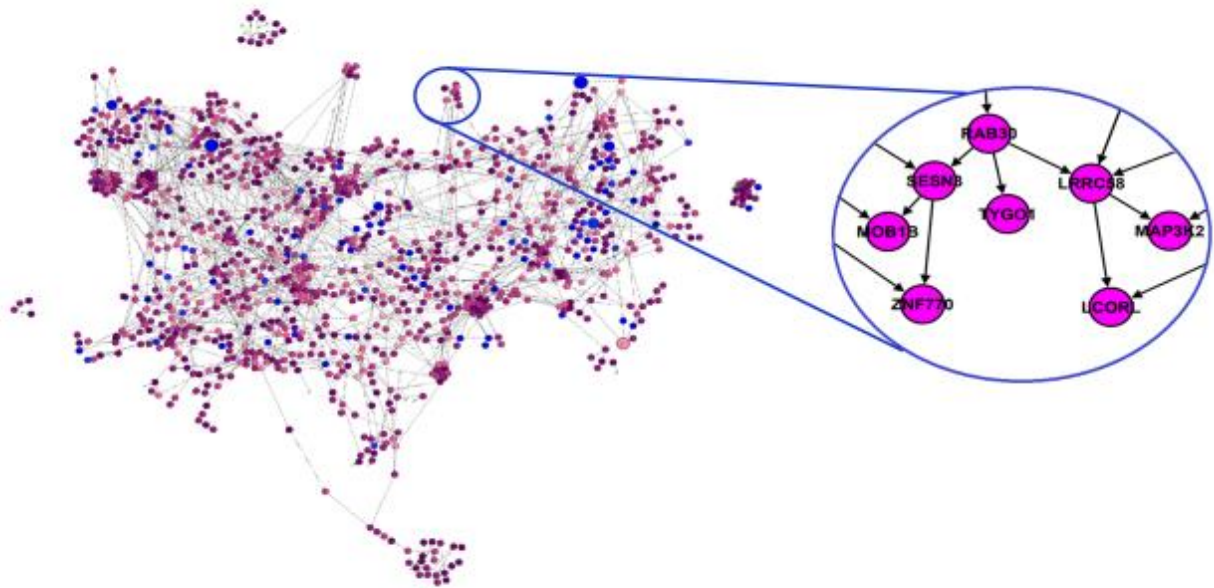

Figure 3. Left: Identified transcriptomic causal network. Each node stands for a gene. Genes associated with SCZ are depicted in blue. Right: A zoom out relationships among some genes in the network.

**High impact genes**: Some genes relative to others have high impact in the system (directly/indirectly) due to having a high out-degree and a high number of maximum effect-blocking-step (Yazdani *et al.*, 2016b). Out degree refers to the number of genes directly affected by a gene. It represents the impact of a gene in the system. For instance, *RAB30* in Figure 3, right panel, has 3 out-degree. The effect of a gene can be propagated serially, until reaching a blocking gene which is a gene with zero out-degree. In Figure 3 right panel, *ZNF770*, *LCOR*, *MAP3K2* and *MOB1B* are the blocking genes. The effect blocking step of a gene is defined as the number of genes influenced by a gene in one path before the effect gets blocked. *SENS3* and *LRRC58* are identified with the effect blocking step equal to one in all paths in the right panel. The maximum number of effect blocking steps across all the paths of a gene is called the max effect blocking step, which is 2 for *RAB30*. The low max effect blocking step of a gene represents that the indirect impact of the gene on the transcriptomic network is not high.

Some of the properties of the identified transcriptomic network are summarized in Figure 4 by focusing on SCZ associated genes. The figure on the left side, summarizes the impact of SCZ associated genes in the system. The figure on the right side clusters the SCZ associated genes based on type of studies that previously found the association.

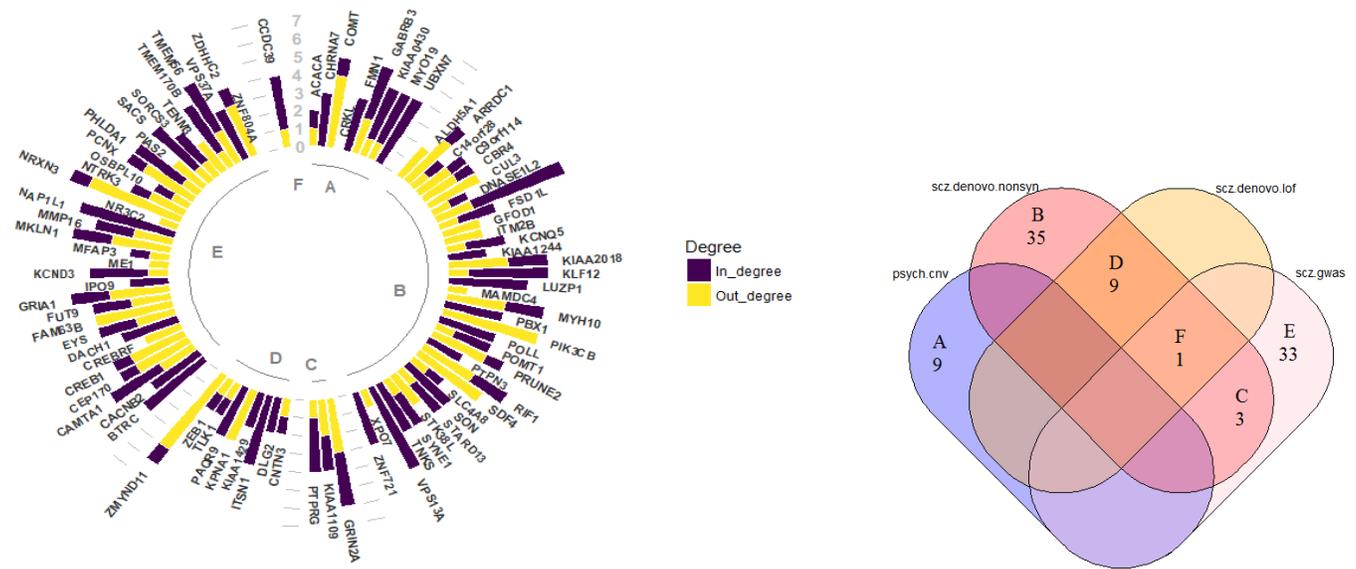

Figure 4. Left: represents the in-degree and out-degrees of genes associated with SCZ in the network. Right: Vann diagram cluster those genes based on the type of studies that previously found their association with SCZ.

**Modules:** A module includes genes influenced by a high impact gene directly or indirectly. We identified several modules across the transcriptomic network with a focus on SCZ associated genes summarized in Table 1. Through further analyses of the modules, we discussed the gene-gene underlying relationships (pathways) to identify novel SCZ associated genes and intervention targets summarized in Table 2. For each module, a heatmap is provided to present the strength of the pathways based on the p-values' cut-off. Moreover, we analyzed the gene ontology for each module represented below. Due to space limitation, the modules are depicted as a sub-network.

Table 1 summarized the main features of the identified modules; name of the hub, number of genes in the module, the number of genes influenced by the hub in the longest path (Max effect blocking step), the number of genes associated to SCZ and the gene ontology.

| Module | SCZ | #genes | # SCZ associated genes | Max effect blocking step | Gene ontology |
|---|---|---|---|---|---|
| NRXN3 | SCZ-GWAS | 9 | 3 | 2 | Gene Expression, RNA damage & repair, RNA post-transcriptional modification |
| COMT | Psych-cnv | 12 | 2 | 2 | Drug Metabolism, Molecular Transport, Cell-To-Cell Signaling & Interaction |
| PEX5L | ----- | 10 | 2 | 4 | Cellular development, cellular Growth & proliferation, nervous system development & function |
| MYH10 | SCZ-denovo-nonsyn | 10 | 2 | 3 | cellular development, cellular growth & proliferation, nervous system development & function |
| TENM3 | ----- | 10 | 5 | 4 | cellular Movement, reproductive system development & function, cardiovascular system development & function |

Table 2. hypotheses made through further analysis of the transcriptomic causal network.

| Hypotheses | | | |
|---|---|---|---|
| SCZ associated genes | | Intervention target for SCZ associated genes | |
| Genes | Modules | Target | SCZ associated genes |
| *GRIN3A* | *NRXN3*-Module | *MAS1* | *NRXN3* |
| *RNF150* | *NRXN3*-Module | *PEX5L* | *STARD13* |
| ANK2 | *MYH10*-Module | *PEX5L* | *DLG2* |
| *GRIN3A* | *TENM3*-Module | | |

*NRXN3*-**Module:** NRXN3 as the hub of the identified transcriptomic modules encodes a member of a family of proteins that function in the nervous system as receptors and cell adhesion molecules. *NRXN3* is reported as SCZ associated gene (Hu et al., 2013) through genome wide association study. The module includes 6 genes influenced by *NRXN3* directly (*ARFGEF3*, *HECW2*, *SORCS3*, *DLGAP1*, *CNTNAP2*) and indirectly (*RNF150*) while its effects are blocked in the module after one or two steps.

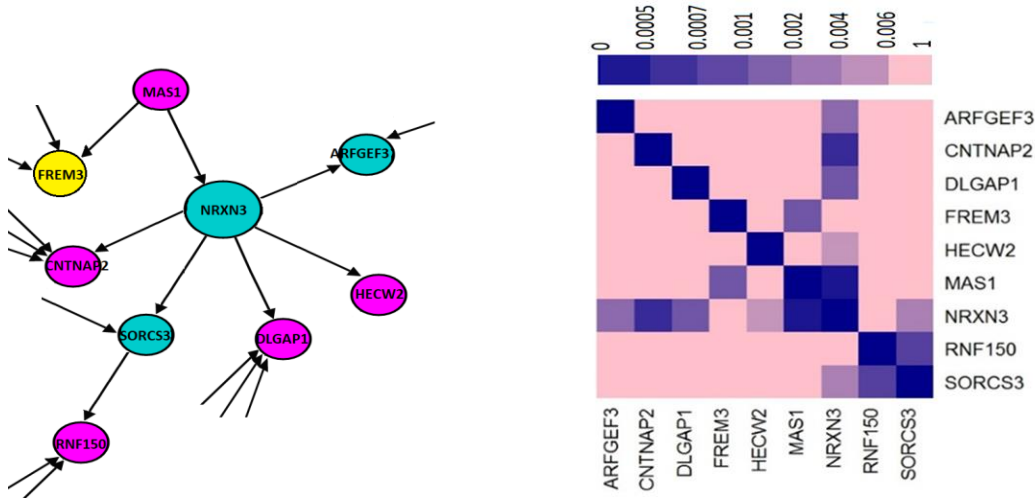

Figure 5. Left: *NRXN3*-Module with multiple SCZ associated genes depicted in blue. Right: heatmap of the strength of the pathways based on the p-values' cut-off.

As Figure 5 shows, the longest path in this module is *MAS1*→ *NRXN3*→ *SORCS3*→ *RNF150* which includes two genes associated with SCZ presented in blue. Therefore, we hypothesize that *RNF150* is probably associated to SCZ as well. The strongest pathway in this module is *MAS1*→ *NRXN3*→ *CNTNAP2*.

The network reveals that *MAS1* has impact on *NRXN3* and *FREM3* while the impact on *NRXN3* is much higher. NRXN3 distributes the effect of MAS1 into the system although FRM3 blocks the effect of MAS1. Interestingly, the gene ontology analysis shows that all genes in the module except FRM3 have similar function related to "gene expression, RNA damage and repair, RNA post transcriptional and modification". FRM3 that blocks the effect of MAS1 does not have the same functionality as MAS1 and its functionality is related to "cellular Growth and proliferation, organismal development and cell cycle". Therefore, MAS1 is a good target for intervention since it is the cause of SCZ associated gene, NRXN3, with high impact and similar function.

**COMT-Module:** Catechol-O-methyltransferase (*COMT*) is one of the hubs with impact on 10 genes in the identified transcriptomic network. Figure 6 shows underlying relationships among these genes where the effect of *COMT* is blocked after at most two steps. The strongest path in the module is through *DCXR* (*COMT*→ *DCXR*→ *YIF1A*).

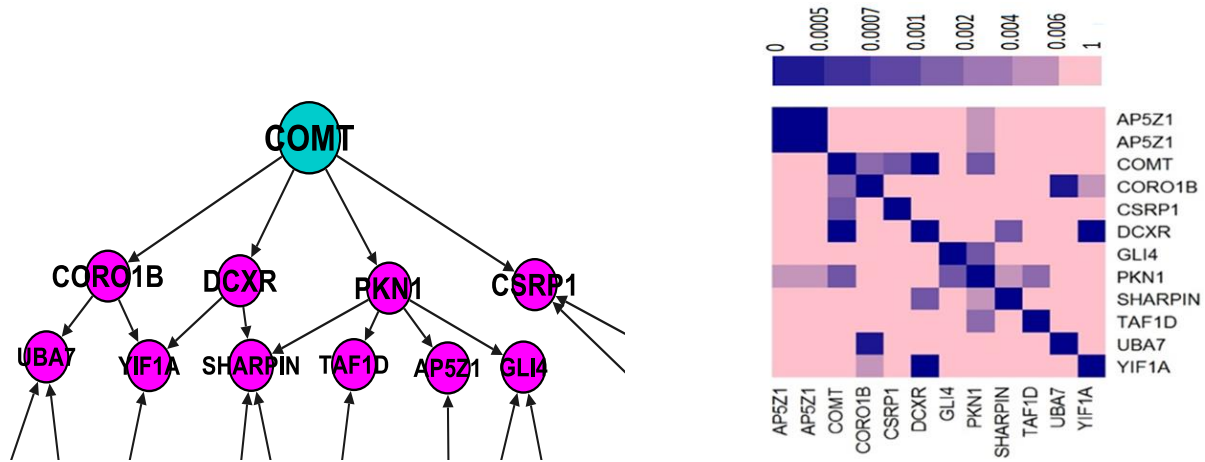

Figure 6. *COMT*- Module with a hub associated with SCZ in blue. Right: heatmap of the strength of the pathways based on the p-values' cut-off.

Based on gene ontology analysis, the *COMT*-Module is related to "Drug Metabolism, Molecular Transport, Cell-To-Cell Signaling and Interaction". *COMT* plays an essential role in the module by influencing multiple genes from the same functionality based on gene ontology. This information makes *COMT* a good target for intervention, which is discussed with more details in the discussion section.

*GLI4* is the only gene that does not have similar function as the other genes in the module. *GLI4* is a blocking effect gene in the boundary of the module with a moderate effect from *COMT* (see the right panel in Figure 5). As Figure 5 shows, two other genes from outside the module influence *GLI4*. The different functionality of *GLI4* is due to the larger effect size on *GLI4* from outside of module compared to the inside.

*MYH10*-**Module**: *MYH10* is an associated SCZ gene influencing 7 genes in the network while influenced by two genes (*ADAM23* and *CHD6*). The longest path that spreads the effect of *MYH10* in the system is *CHD6*→*MYH10*→*LMO7*→*RTF1*→*HMG20A*. This path is also the strongest path in this module.

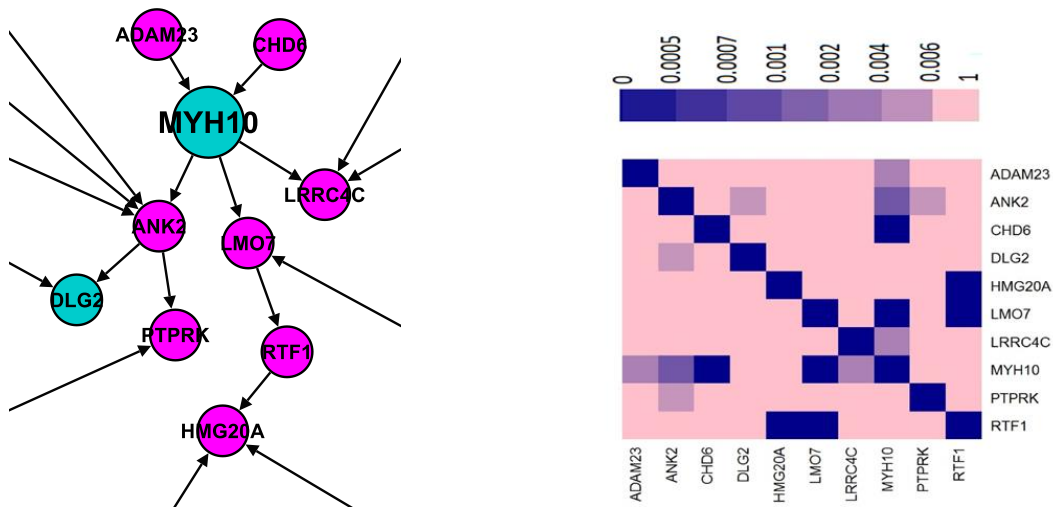

Figure 7. *MYH10*- Module with a hub associated with SCZ in blue. Right: heatmap of the strength of the pathways based on the p-values' cut-off.

Figure 7 shows a path of three genes (*MYH10*→*ANK2*→ *DLG2)* while two of these genes are associated with SCZ identified via *de novo* LoF mutation studies. Therefore, we hypothesize that *ANK2* is potentially associated with SCZ. Gene ontology analysis categorized all genes in *MKLN1* module with similar Function, "cellular development, cellular growth and proliferation, nervous system development and function".

***PEX5L*-Module**: *PEX5L* is the hub of this module with significant impacts on two genes associated with SCZ (*STARD13* and *DLG2*). The strong relationship between *PEX5L* and these two genes make *PEX5L* a good target for intervention in the study of SCZ. Considering the underlying relationship of *PEX5L* provides insights on designing for the intervention and better understanding of its effect in the system.

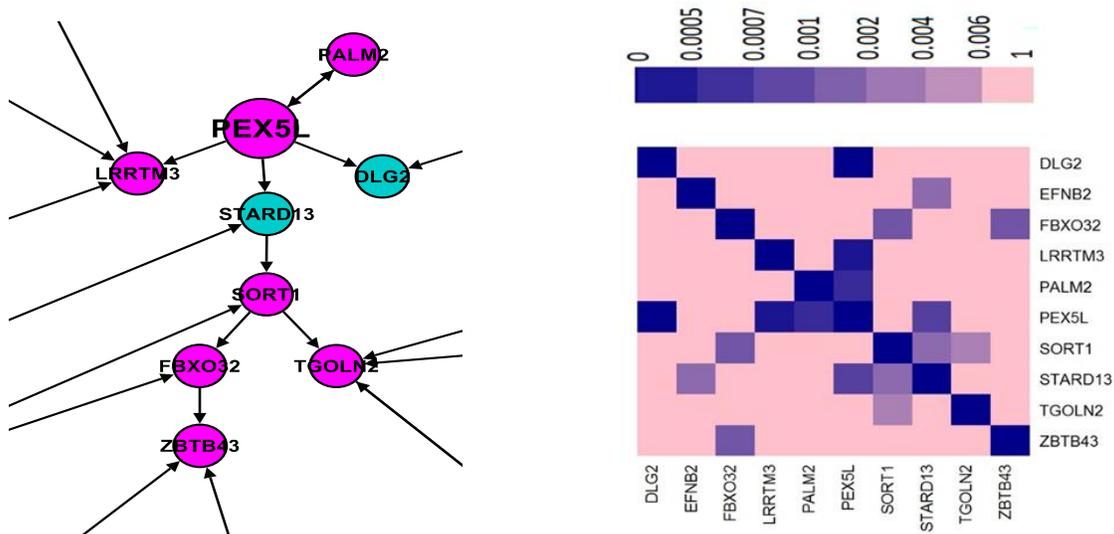

Figure 8. *PEX5L*-Module with a hub associated with SCZ in blue. Right: heatmap of the strength of the pathways based on the p-values' cut-off.

As Figure 8 shows, the direction of *PALM2* is not identified through this analysis. While *PALM2* is in the edge of the network and no IVs affected on this gene, we hypothesize that *PALM2* may be affected by the genes that are not included in this study. Gene ontology shows all genes in *PEX5L*-Module including *PALM2* have similar function related to "Cellular development, cellular Growth and proliferation, nervous system development and function".

***TENM3*-Module:** *TENM3* is a SCZ associated gene with impact on 10 genes in the network while 4 of them are associated with SCZ. The longest path in this network is *TENM3*→ *GRIN3A*→ *GRIA1*→ *SORCS3*→ *RNF150*. In this pathway, all the genes are associated with SCZ except *GRIN3A* and *RNF150*. We can make a strong hypothesis that *GRIN3A* is an SCZ associated gene since it is influenced only by one gene which is associated to SCZ and *GRIN3A* itself influences 3 genes associated with SCZ directly or indirectly. RNF150 may also be associated with SCZ since it strongly influenced by SORCS3 and not by SEZ6L.

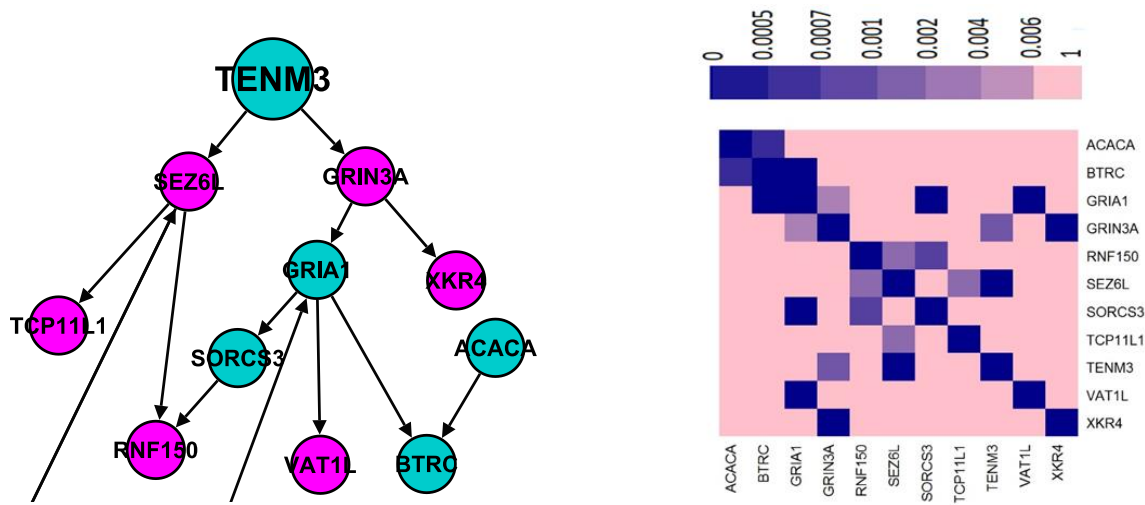

Figure 9. *TENM3*- Module with a hub associated with SCZ in blue. Right: heatmap of the strength of the pathways based on the p-values' cut-off.

Gene ontology shows that all genes in this module are related to functionality "cellular Movement, reproductive system development and function, cardiovascular system development and function".

**Discussion**

Revealing the underlying relationships among genes is complementary to experimental study through facilitating mechanistic understanding and providing insights to pathology of disease and treatment. Focusing on SCZ as a heritable complex mental disorder, we selected a cluster of 1181 genes with a large number of schizophrenia associated genes from the study by Fromer, *et al.* (2016). To construct causal network in observational study, we modified the G-DAG algorithm for RNA-seq data, which is an algorithm based on Mendelian randomization and Bayesian graphical modeling. We utilized the fact that genetic variants are the cause of substantial variation in transcriptomic level and therefore, extracted information from 958,178 genetic variants (SNPs) and identified underlying relationships among the transcriptomics. Mapping the SCZ associated genes on the network, we elicited information from the network such as, transcriptomic modules, high impact genes and underlying pathways in the system to provide a mechanistic understanding.

This study provides the first algorithm to reveal underlying relationships among schizophrenia candidate genes in large scale. Our preliminary results suggest that this systems approach is likely to lead to establishing etiological mechanisms of disorder, including psychiatric illness. Our examination of the network showed that the impact of a specific gene is on few genes and does not spread across the system. In addition, we hypothesized some genes associated to the disorder as well as causes of the associated genes and targets for intervention. Furthermore, the underlying gene relationship provided some insights into how designing the intervention.

We hypothesized the gene *MAS1* as an intervention target for the SCZ associated gene *NRXN3*. The revealed pathways provide some insights that how the effect of an intervention in the gene *MAS1* spreads through the system and how disease mechanism may break down these pathways. Analyses of *TENM3* module shows that *GRIN3A* directly and indirectly influences 3 genes associated with SCZ. Furthermore, *GRIN3A* is influenced by only one gene which is associated to SCZ too. This information hypothesized that *GRIN3A* itself is also a SCZ associated gene.

*COMT* was identified as a good target for intervention since it is not influenced by any gene and influenced multiple genes from the same functionality, based on gene ontology. *COMT* is an important regulator of a number of cognitive and behavioral processes due to its role in the termination of dopamine signaling in brain regions such as the prefrontal cortex and hippocampus. There are many studies (e.g. Saravani, 2017 ; Bhakta, 2012) on brain-penetrating inhibitors of *COMT* for curing a variety of conditions associated with dysregulated cortical dopaminergic function like schizophrenia. However, the known brain penetrant *COMT* inhibitors nitrocatechol tolcapone has serious safety issues that preclude widespread use in psychiatric disorders. Therefore, understanding *COMT* systematic effect on other genes provides insights into its functional mechanism in the brain and is useful for treatment of neurologic disorders such as SCZ (Barrow and Akuma, 2017).

Identifying underlying pathways among genes reveals important biological relationships among molecular components and leads to understanding complex biological mechanisms, disease pathogenesis, and underlying co-morbidities (Zhernakova, Van Diemen and Wijmenga, 2009; Lee *et al.*, 2012). This understanding may improve and speed drug development and furthermore, predict possible side effects (van de Peppel and Holstege, 2005; Williams, 1957). Although systems approaches to molecular pathway identification will not replace mechanistic experiments, they are complementary and hypothesis-generating especially in the age of large scale data to narrow the search space, identify targets for intervention, and follow the effect of any intervention in the system.

**Acknowledgments**

The first author thanks Dr. Panoss Rousoss for introducing us the data and providing comments. We thank CommonMind Consortium for providing the data.

**Conflict of interest**

There is no conflict of interest among authors.